\documentstyle[preprint,aps,psfig]{revtex}
\begin{document}
\draft
\preprint{MKPH-T-96-9}
\title{Rescattering effects in coherent pion photoproduction on the
  deuteron in the $\Delta$ resonance region\thanks{Supported by the
    Deutsche Forschungsgemeinschaft (SFB 201)}} \author{P.\ Wilhelm
  and H.\ Arenh\"ovel} \address {Institut f\"ur Kernphysik, Universit\"at
  Mainz, D-55099 Mainz, Germany} 
\maketitle
\begin{abstract}
  Within a dynamical model, rescattering effects are studied in
  coherent pion photoproduction on the deuteron in the $\Delta$(1232)
  resonance region.  The model consists of a system of coupled
  equations for the N$\Delta$, NN$\pi$ and NN channels.  Pion
  rescattering leads in general to a significant reduction of total
  and differential cross sections resulting in a better description of
  experimental data.  Only a few polarization observables are
  sensitive to rescattering effects.  In particular $T_{20}$ allows an
  analysis of the reaction mechanism for specific kinematics.  The
  question how rescattering affects the sensitivity to the E2
  excitation of the $\Delta$ resonance is also studied.
\end{abstract}
\pacs{PACS numbers: 13.60.Le, 21.45.+v, 25.20.Lj}
\narrowtext
\newpage

\section{Introduction}

Recently, we have studied coherent pion photoproduction on the
deuteron neglecting rescattering effects in \cite{WA1}, henceforth
referred to as I.  Our justification for this work was to study
systematically the details of the elementary producton amplitude on
the one hand, and on the other, to investigate the influence of
genuine two-nucleon mechanisms which had not been considered before.
However, the comparison of theoretical predictions and experimental
data in I as well as in previous studies (see e.g.\ \cite{bbb1}) gave
clear indication of important rescattering effects.

In most of former work, see e.g.\ \cite{OsR76,LlM76,BoL78}, the
rescattering of the pion was taken into account only perturbatively.
For a recent study including single- and double-scattering diagrams,
we refer to the work of Garcilazo and Moya de Guerra \cite{garcilazo}.
It is known that neutral pion production for photon energies just
above threshold up to, say 450$\,$MeV, is dominated by intermediate
$\Delta$ resonance excitation via an electromagnetic M1 transition and
subsequent $p$ wave pion emission.  However, it seems rather
questionable whether a strong interacting system like the N$\Delta$
system in a relative $^5S_2$ partial wave with vanishing angular
momentum barrier can be treated perturbatively at all.  Thus a full
dynamical treatment of at least the N$\Delta$ sytem appears
mandatory, and it is the aim of this paper to develop such a
description.

Only recently, different non-perturbative treatments of pion
rescattering in $\gamma d\rightarrow \pi^0 d$ became available which
can be applied in the $\Delta$ region.  Blaazer, Bakker and Boersma
\cite{bbb} used the Faddeev equations for the $\pi$NN system.  One of
their main conclusions was that only the rescattering on the
single-nucleon level is significant whereas genuine three-body effects
turn out to be fairly unimportant in the $\Delta$ region. In a very
recent work, Kamalov, Tiator and Bennhold applied the KMT multiple
scattering approach \cite{kmt} to coherent pion photoproduction
\cite{ktb} as well as to elastic pion scattering on the deuteron
\cite{ktbpi}.  A characteristic feature of their approach is that it
neglects the contributions from the coupling with the break-up
channels.  We will discuss below in some detail the shortcomings of
this assumption.  A dynamical treatment of the $\Delta$,
conceptionally similar to the present work, has been used by Pe\~{n}a
et al.\ \cite{pena}.  However, the main point of their paper was not
the $\gamma d\rightarrow \pi^0 d$ reaction itself, and therefore, the
presentation of their results is very brief.  Moreover,
the treatment of the nonresonant amplitudes and the input on the
electromagnetic side is not clear.

Despite all these efforts, no common conclusions as to the importance
of rescattering effects is reached.  In fact, the various approaches
even disagree qualitatively.  For example at 300$\,$MeV photon energy,
the rescattering amplitudes of Ref.\ \cite{ktb} within the KMT
approach reduce the cross section. Their influence grows with
increasing pion angle up to a reduction by a factor three at
180$\,$deg.  Also within the Faddeev ansatz of Ref.\ \cite{bbb}, the
cross section is reduced in the forward direction by rescattering but
at backward angles it is slightly increased.  Unfortunately, the
results shown in Ref.\ \cite{pena} do not allow to separate the rescattering
influence.  Thus the role of pion rescattering in $\gamma
d\rightarrow\pi^0 d$ is by no means settled and requires further
investigation.

This paper is organized as follows.  In Section \ref{sec2} we present
our dynamical model for the N$\Delta$ system.  As a test of the
model, cross sections for pion deuteron elastic scattering are shown
in Section \ref{sec3}.  The rescattering contributions which follow
from the application of the model to coherent pion photoproduction on
the deuteron are given in Section \ref{sec4}.  The rescattering of
charged pions which have not been produced by an initial
electromagnetic excitation of the $\Delta$ is added in an approximate
way.  Our results for $\gamma d\rightarrow \pi^0 d$ are presented in
Section \ref{sec5}.  Some emphasis is put on the role of charge
exchange in the rescattering process.  In our study of polarization
observables we focus on the tensor target asymmetry $T_{20}$ which may
help to disentangle different reaction mechanisms.  Finally, we study
the sensitivity of the vector target asymmetry $T_{11}$ to the E2
excitation of the $\Delta$ resonance.

\section{Dynamical model for the $N\!\Delta$ system}
\label{sec2}

The theoretical concept of our model with respect to the hadronic part
is mainly based on the developments of Sauer and collaborators
\cite{ha1,ha2} and is similar to the treatment of Lee et al.\ in Ref.\ 
\cite{lee1,lee2,lee3,lee4}.  The model includes explicit pion, nucleon
and delta degrees of freedom.  Hence its configuration space $\cal{H}$
is built up from NN-, N$\Delta$- and $\pi$NN-sectors
\begin{equation}
\cal{H}=\cal{H}_{NN}\oplus\cal{H}_{N\Delta}\oplus\cal{H}_{\pi NN}.
\end{equation}
Corresponding projectors are denoted as $P_N$, $P_{\Delta}$ and $Q$,
and we use for any operator $\Omega$ the obvious notation
$\Omega_{N\Delta}=P_N\Omega P_{\Delta}$, $\Omega_{\Delta
  Q}=P_{\Delta}\Omega Q$, etc.  The hamiltonian, $H=H_0+V$, contains
the kinetic energy $H_0$ and an interaction $V$ which can be split up
according to the various sectors
\begin{equation}
  V = \sum_{X,Y=N,\Delta,Q}V_{XY}.
\end{equation}

In the spirit of this model, we assume $V_{NQ}=V_{QN}=0$, i.e., there
is no explicit $\pi$NN-vertex leading to a direct coupling between
NN- and $\pi$NN-sector.  The creation or annihilation of a pion is
exclusively due to the $\pi$N$\Delta$ vertex $v_\Delta$ already
introduced in (I,19). It defines $V_{\Delta Q}=v_\Delta(1) +
v_\Delta(2)$ and $V_{Q\Delta}=v^\dagger_\Delta(1) +
v^\dagger_\Delta(2)$.  As a further approximation, we switch off the
diagonal interaction inside the $\pi$NN-sector completely, i.e.,
$V_{QQ}=0$.  With the latter simplification, the problem is
essentially reduced to an effective two-body one.  This holds
irrespective of the specific form assumed for the remaining
interactions $V_{\Delta N}$, $V_{N\Delta}$, $V_{NN}$, and
$V_{\Delta\Delta}$.  All dynamics is contained in a transition
amplitude $T$ which obeys a coupled integral equation of the
Lippmann-Schwinger type
\begin{eqnarray}
  && T_{NN}(W^+) = V_{NN} + V_{NN} G_N(W^+) T_{NN}(W^+) + V_{N\Delta}
  G_\Delta(W^+) T_{\Delta N}(W^+),
    \label{eq:cc}\\
    && T_{\Delta N}(W^+) = V_{\Delta N} + V_{\Delta N} G_N(W^+)
    T_{NN}(W^+) +V_{\Delta \Delta}^{eff}(W^+) G_\Delta(W^+) T_{\Delta
      N}(W^+), \nonumber\\ 
    && T_{N \Delta}(W^+) = V_{N \Delta} + V_{NN} G_N(W^+)
    T_{N\Delta}(W^+) +V_{N \Delta}(W^+) G_\Delta(W^+)
    T_{\Delta\Delta}(W^+), \nonumber\\ 
    && T_{\Delta \Delta}(W^+) = V_{\Delta \Delta}^{eff}(W^+) +
    V_{\Delta N} G_N(W^+) T_{N \Delta}(W^+) +V_{\Delta
      \Delta}^{eff}(W^+) G_\Delta(W^+) T_{\Delta \Delta}(W^+),
    \nonumber
\end{eqnarray}
where $W$ is the invariant mass of the system.  In (\ref{eq:cc}),
$G_\Delta$ denotes the dressed propagator of the 
noninteracting N$\Delta$ states,
already introduced in (I,41-42), and
\begin{equation}
G_N(W^+)=\left[W+i\epsilon-h_N(1)-h_N(2) \right]^{-1} 
\end{equation}
is the free propagator in the NN-sector with the free nucleon
energy $h_N=M_N+\vec p^{\,2}/(2M_N)$.

For the present application, the most relevant part of the driving
term in (\ref{eq:cc}) is $V^{eff}_{\Delta\Delta}(z)$ which describes the
N-$\Delta$ interaction.  It contains a retarded (ret) and thus
energy-dependent pion exchange $V_{\Delta\Delta}^{ret}(z)$ and a
static part which is just given by $V_{\Delta\Delta}$,
\begin{equation}
V^{eff}_{\Delta\Delta}(z)=V_{\Delta\Delta}^{ret}(z) + V_{\Delta\Delta}.
\end{equation}
The first one follows from the iteration of the $\pi N\Delta$ vertex
leading to
\begin{equation}
  V_{\Delta\Delta}^{ret}(z)=\left( V_{\Delta Q} \frac{1}{z-H_0}
    V_{Q\Delta} \right)_{[2]}\,,
\label{eq:vdd}
\end{equation}
where the subscript $[2]$ indicates the restriction to the two-body
part of the operator.  Besides the propagator $G_\Delta$, this
effective potential $V_{\Delta\Delta}^{ret}(z)$ contains implicitly a
coupling to the three-body sector as required by three-body unitarity.
Furthermore, the above mentioned neglect of an interaction in the
$\pi$NN space ($V_{QQ}=0$) leads to a partial violation of two-body
unitarity since the coupling to intermediate $\pi d$ states is not
included.

In detail we obtain for (\ref{eq:vdd})
\begin{equation}
  V_{\Delta\Delta}^{ret}(z)= \vec\tau_{\Delta N}(1) \cdot
  \vec\tau_{N\Delta}(2) \int \frac{d^3q}{(2\pi)^3 2\omega_{\vec q}}
  e^{i\vec q\cdot\vec r} \frac{\vec{\sigma}_{\Delta N}(1) \cdot \vec{q} \,
    \vec{\sigma}_{N\Delta}(2) \cdot \vec{q}} {z-\omega_{\vec
      q}-h_N(1)-h_N(2)} \frac{F_\Delta^2(\vec q^{\,2})}{m_\pi^2}
  +(1\leftrightarrow 2),
\label{eq:ret}
\end{equation}
where $\omega_{\vec q}=(m_\pi^2+\vec q^{\,2})^{1/2}$. The $\pi$N$\Delta$
form factor $F_\Delta$ is given in (I,22) and the normalization of the
spin (isospin) operators $\vec\sigma_{N\Delta}$ ($\vec\tau_{N\Delta}$)
in (I,21).  In order to arrive at (\ref{eq:ret}) we have dropped the
transformation (I,20), i.e., we have replaced $\vec q_{\pi N}$ by
$\vec q$ in the $\pi$N$\Delta$ vertex (I,19).

For the static potential $V_{\Delta\Delta}$, we consider only pion
exchange associated with intermediate $\pi\Delta\Delta$ states and
neglect heavier meson contributions.  Thus $V_{\Delta\Delta}$ follows
from (\ref{eq:ret}) by setting $z-h_N(1)-h_N(2)=0$ in the denominator.
In addition, $V$ contains a NN$\leftrightarrow$N$\Delta$ transition
potential $V_{N\Delta}=V_{\Delta N}^\dagger$ which is given in
Appendix B and a diagonal NN interaction $V_{NN}$ which is based
on the OBEPR parametrization of the Bonn potential \cite{machleidt}.
In view of the fact that the coupling to intermediate two-nucleon
states turns out to be very small for coherent pion photoproduction we
give no details on the construction of $V_{NN}$ here and refer to
\cite{wilhelm} for the specification of this part of the potential
where the model was applied to two-body deuteron photodisintegration.

For the explicit calculation we have solved Eq.\ (\ref{eq:cc})
numerically in momentum space.  Some details are given in Appendix A.

\section{Pion deuteron elastic scattering as a test of the model}
\label{sec3}

\begin{figure}
  \centerline{\psfig{figure=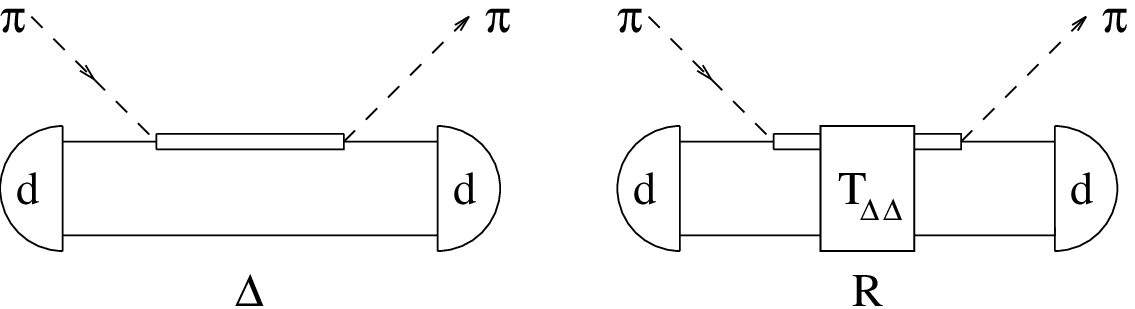,width=10cm,angle=0}}
\caption{
  Representation of the $\pi d\rightarrow\pi d$ amplitude including a
  direct ($\Delta$) and a rescattering amplitude (R).}
\label{fig:pidamplitudes}
\end{figure}

\begin{figure}
  \centerline{\psfig{figure=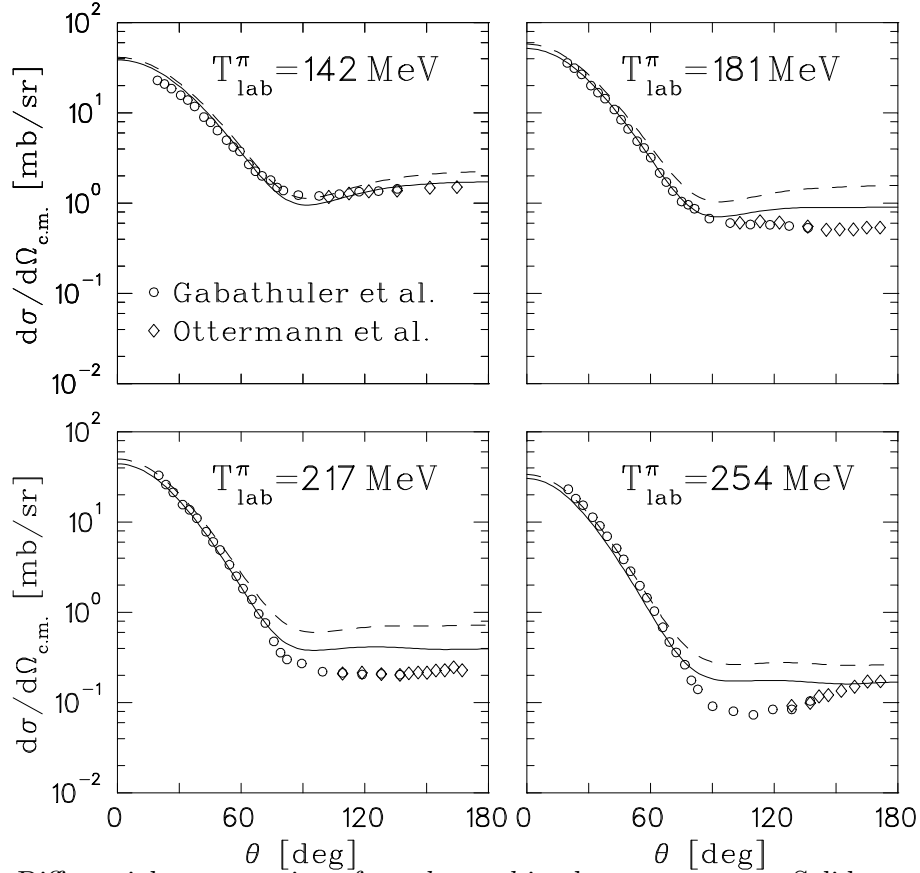,width=12cm,angle=0}}
\caption{
  Differential cross sections for $\pi d\rightarrow\pi d$ in the c.m.\ 
  system. Solid curves: complete calculation, and dashed curves:
  without rescattering (R). Experimental data from
  \protect\cite{gabathuler} (circles) and \protect\cite{ottermann}
  (rhombs).}
\label{fig:pid}
\end{figure}

Before we apply the hadronic interaction model to coherent pion
photoproduction, we will first consider elastic pion deuteron
scattering as a test case. Within our model, the amplitude for this
process consists of a direct ($\Delta$) and a rescattering (R) term, $
T_{\pi d}=T_{\pi d}^{(\Delta)}+T_{\pi d}^{(R)}$, corresponding to the
diagrams in Fig.\ \ref{fig:pidamplitudes} with
\begin{eqnarray}
  \label{eq:pid}
  T_{\pi d}^{(\Delta)}(W_{\pi d}^+) &=& V_{Q\Delta} G_\Delta(W_{\pi d}^+)
  V_{\Delta Q},\\ 
  T_{\pi d}^{(R)}(W_{\pi d}^+) &=& V_{Q\Delta} G_\Delta(W_{\pi d}^+)
  T_{\Delta\Delta}(W_{\pi d}^+) G_\Delta(W_{\pi d}^+) V_{\Delta Q},
\end{eqnarray}
where $W_{\pi d}$ is the invariant mass of the process.  We show in
Fig.\ \ref{fig:pid} the resulting differential cross sections for
various pion energies.  Apparently, the rescattering contribution,
which reduces the cross section up to a factor two at backward angles,
leads to a considerably better description of the data in the
resonance region ($W_{\pi d}=M_N+M_\Delta$ corresponds to the kinetic
pion laboratory energy $T^{\pi}_{lab}\approx 180\,$MeV) although at
backward angles some overestimation remains.  From this we conclude
that the model presents a reasonable starting point to study pion
rescattering effects in $\gamma d\rightarrow\pi^0 d$.

\section{Application to coherent pion photoproduction}
\label{sec4}

\begin{figure}
  \centerline{\psfig{figure=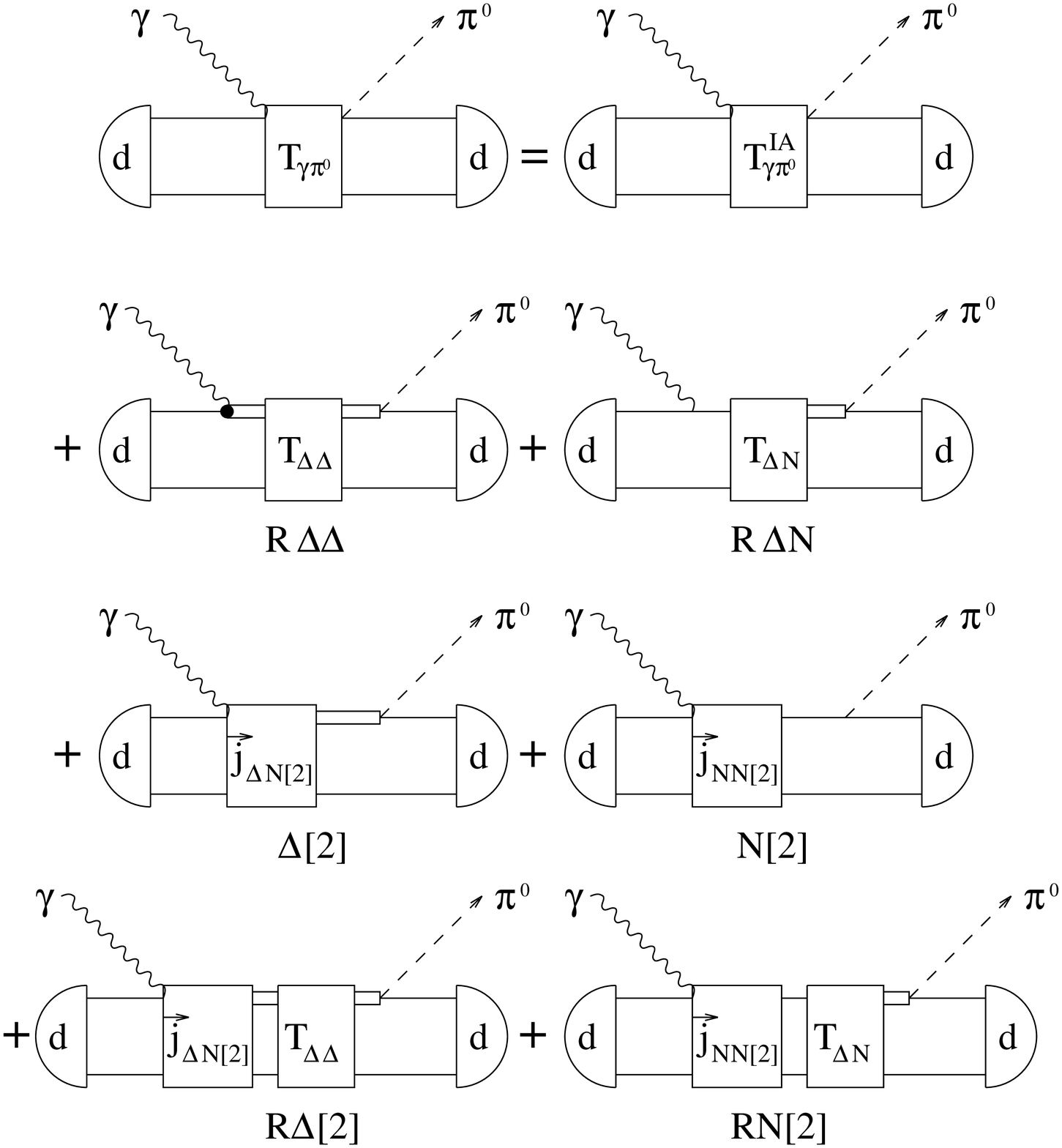,width=12cm,angle=0}}
\caption{
  The complete amplitude for $\gamma d\rightarrow\pi^0 d$ of the
  present model contains the direct contributions
  $T_{\gamma\pi^0}^{IA}$ defined in I and shown in Fig.\ 
  \protect\ref{fig:amplitudes1} and rescattering contributions.  The
  latter are the graphs generated by the hadronic $T$-matrices
  R$\Delta\Delta$ and R$\Delta$N, the MEC graphs $\Delta$[2] and N[2],
  and the graphs which combine MECs and $T$-matrices R$\Delta$[2] and
  RN[2].  The short-hand notation R=\{R$\Delta\Delta$, R$\Delta$N,
  R$\Delta$[2], RN[2]\} is used in the text.}
\label{fig:amplitudes}
\end{figure}

\begin{figure}
  \centerline{\psfig{figure=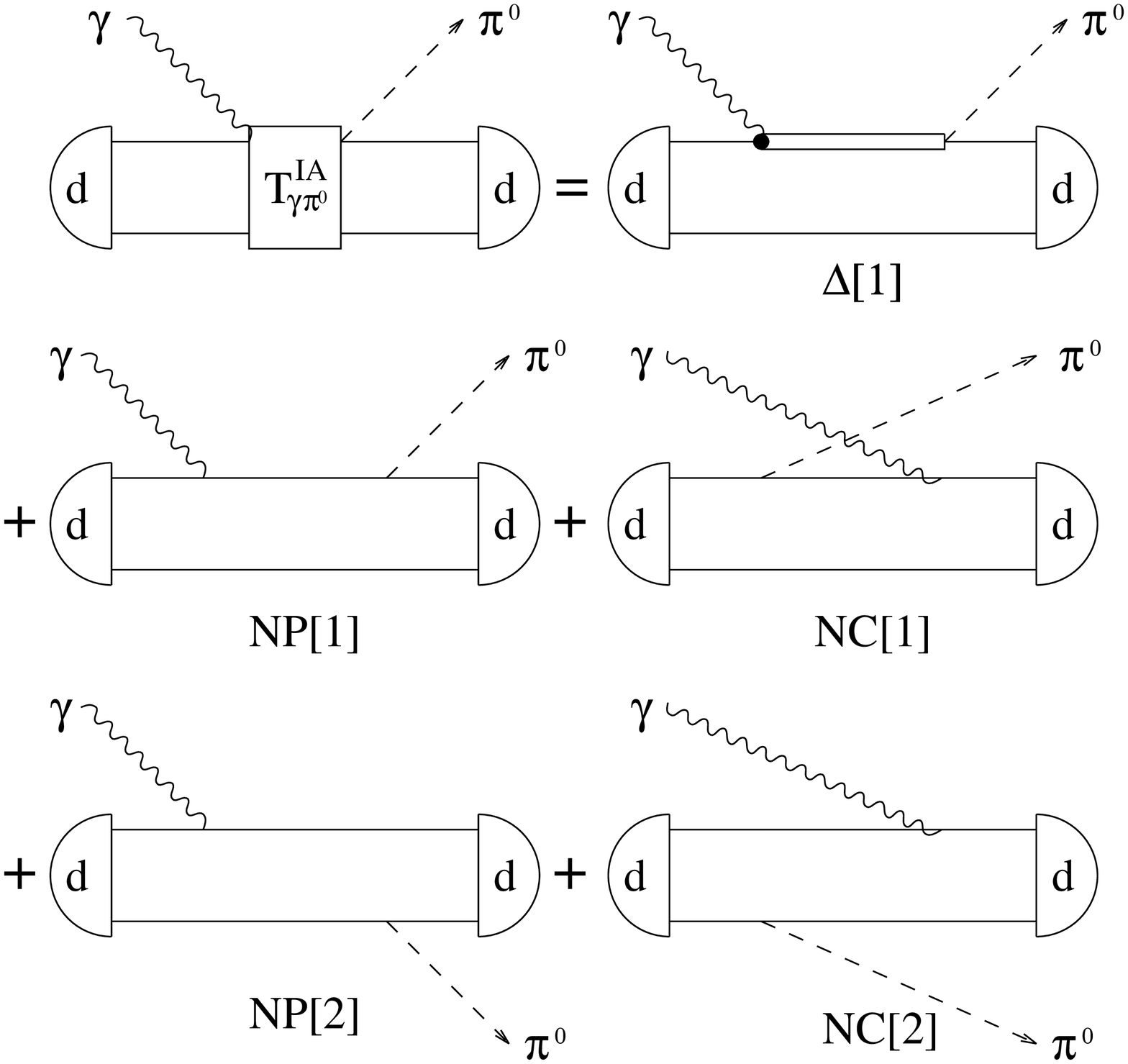,width=9cm,angle=0}}
\caption{
  Contributions to the $\gamma d\rightarrow\pi^0 d$ amplitude
  as studied in I: direct $\Delta$ contribution $\Delta$[1], direct
  NP[1] and crossed NC[1] nucleon pole, direct NP[2] and crossed
  NC[2] disconnected two-body processes.}
\label{fig:amplitudes1}
\end{figure}

Now we will turn to coherent pion photoproduction on the deuteron in
order to study the rescattering effects generated by our hadronic
interaction model.  We show in Fig.\ \ref{fig:amplitudes} explicitly
all diagrams of the process included in this work which go beyond the
ones considered in I. The latter are summarized by
$T_{\gamma\pi^0}^{IA}$ and shown in Fig.\ \ref{fig:amplitudes1}
separately for completeness.  They include the impulse approximation
diagrams, namely the direct $\Delta$ excitation $\Delta$[1], the
direct and crossed nucleon pole terms NP[1] and NC[1], respectively,
and the disconnected direct and crossed two-nucleon processes, NP[2] and
NC[2], respectively.  Formally, the graphs NP and NC follow from the
$\pi$NN vertex $v^\dagger_N$ defined in (I,9).  This explicit vertex,
which defines $V_{QN}=v^\dagger_N(1)+v^\dagger_N(2)$ and
$V_{NQ}=v_N(1)+v_N(2)$, is however treated in first order only.

The common feature of the diagrams in Fig.\ \ref{fig:amplitudes1} is
that the hadronic intermediate state, either $\Delta$N, NN or $\pi$NN, 
propagates freely.  In principle each diagram will be accompanied
by a corresponding one, where the intermediate state is subject to the
hadronic interaction. Furthermore, the interaction will allow
couplings between different intermediate states.  In this work,
however, we will restrict the interaction to the intermediate
N$\Delta$ state resulting in the rescattering amplitude
R$\Delta\Delta$ and to the
NN-N$\Delta$ coupling R$\Delta$N in Fig.\ \ref{fig:amplitudes}.
The contribution of R$\Delta\Delta$ to the
$T$-matrix reads in the notation of I
\begin{equation}
  T_{m'\lambda m}^{R\Delta\Delta}(\vec{q},\vec{k}\,)
  =-\vec{\epsilon}_{\lambda}\cdot \left\langle\vec{q},m'\left|
      V_{Q\Delta}\, G_\Delta(W_{\gamma d}^+)\,
      T_{\Delta\Delta}(W_{\gamma d}^+)\, \vec\jmath_{\Delta
        N[1]}(E_\Delta,\vec k) \right|m\right\rangle,
\label{eq:rd}
\end{equation}
where the one-body $\Delta$ excitation current
$\vec\jmath_{\Delta N[1]}$ has been defined in (I,25-26) and (I,42-43).
In R$\Delta$N the photon is first absorbed by the one-nucleon 
current $\vec\jmath_{NN[1]}$ of (I,10) and (I,50)
with subsequent NN-N$\Delta$ transition via $T_{\Delta N}$. 
Its $T$-matrix contribution is given by
\begin{equation}
  T_{m'\lambda m}^{R\Delta N}(\vec{q},\vec{k}\,)
  =-\vec{\epsilon}_{\lambda}\cdot \left\langle\vec{q},m'\left|
      V_{Q\Delta}\, G_\Delta(W_{\gamma d}^+)\, T_{\Delta N}(W_{\gamma
        d}^+)\, \vec\jmath_{NN[1]}(\vec k) \right|m\right\rangle.
\label{eq:rn}
\end{equation}
However, it turns out that R$\Delta\Delta$ is by far the dominant
process. Therefore, we will not separate the two contributions in the
presentation of our results. In additon to these rescattering
processes, we consider also meson exchange current contributions.

\subsection{Meson exchange currents}

\begin{figure}
  \centerline{\psfig{figure=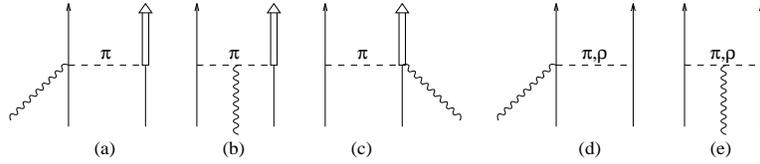,width=10cm,angle=0}}
\caption{
  MEC contributions included in the present calculation.
  $\vec\jmath^{\,\,\pi}_{[2]\Delta N}$: (a)--(c), and
  $\vec\jmath^{\,\,\pi(\rho)}_{[2]NN}$: (d) and (e).}
\label{fig:mec}
\end{figure}

So far we have mainly focused on the rescattering of pions which have
been initially produced via the excitation of the $\Delta$ resonance.
For charged pion production, there are, however, also strong
background contributions, namely, the seagull and the pion pole graph.
The reabsorption on the second nucleon of a pion which has been
produced by one of these mechanisms contributes to the $\pi$-meson
exchange current ($\pi$-MEC).  Strictly speaking, a reabsorption
involving an excitation of a $\Delta$ (Figs.\ \ref{fig:mec}a, b)
corresponds to a subsequent rescattering in the $P_{33}(\pi N)$
channel whereas a reabsorption without $\Delta$ excitation (Figs.\ 
\ref{fig:mec}d, e) contributes to rescattering in the $P_{11}(\pi N)$
channel.  Considering the different time-orderings contained in the
diagrams of Fig.\ \ref{fig:mec}, it is clear, that only the diagrams
with forward propagating pions can be interpreted as a pion
rescattering process.  The graph of Fig.\ \ref{fig:mec}c appears
through minimal coupling and is required by gauge invariance.  In
addition, we also include the $\rho$-MEC. Altogether, the two-body
current $\vec\jmath_{[2]}$ is given by
\begin{equation}
  \label{eq:mec}
  \vec\jmath_{[2]}=\vec\jmath^{\,\,\pi}_{[2]\Delta N}+
  \vec\jmath^{\,\,\pi}_{[2]NN} + \vec\jmath^{\,\,\rho}_{[2]NN}.
\end{equation}
All terms in (\ref{eq:mec}) are evaluated using the static
approximation for the intermediate baryon propagation.  Explicit
expressions are given in Appendix B.  They generate the following
amplitudes including rescattering with $X$ standing for $\Delta$ or N
\begin{eqnarray}
&&
T_{m'\lambda m}^{X[2]}(\vec{q},\vec{k}\,)
=-\vec{\epsilon}_{\lambda}\cdot
\left\langle\vec{q},m'\left|
V_{QX}\, G_X(W_{\gamma d}^+)\, \vec\jmath_{[2]XN}(\vec k)
\right|m\right\rangle,
\\
&&
T_{m'\lambda m}^{RX[2]}(\vec{q},\vec{k}\,)
=-\vec{\epsilon}_{\lambda}\cdot
\left\langle\vec{q},m'\left|
V_{Q\Delta}\, G_\Delta(W_{\gamma d}^+)\, 
T_{\Delta X}(W_{\gamma d}^+)\,
G_X(W_{\gamma d}^+)\, \vec\jmath_{[2]XN}(\vec k)
\right|m\right\rangle.
\label{eq:rmec}
\end{eqnarray}
They are represented in Fig.\ \ref{fig:amplitudes} by the 
diagrams $\Delta$[2], N[2], R$\Delta$[2], and RN[2].

\section{Results for coherent pion photoproduction}
\label{sec5}

Now we will discuss our results for coherent pion photoproduction on
the deuteron.  For the explicit calculation we have used a partial
wave decomposition.  Sufficient convergence is achieved by calculating
the contributions NP[1], NP[2], NC[1] and NC[2] of Fig.\ 
\ref{fig:amplitudes1} up to a total angular momentum $j_{max}=5$, and
$\Delta$[1] up to $j_{max}=10$.  The rescattering graphs R,
$\Delta$[2] and N[2] of Fig.\ \ref{fig:amplitudes} are included up to
$j_{max}=3$.

\subsection{Cross section}

\begin{figure}
  \centerline{\psfig{figure=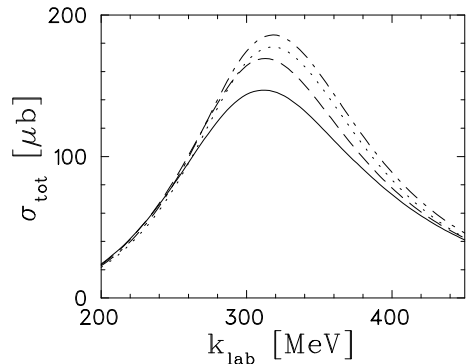,width=6cm,angle=0}}
\caption{
  Total cross section for $\gamma d\rightarrow\pi^0 d$. Solid curve:
  complete calculation, dotted curve: without rescattering amplitudes
  (R, $\Delta$[2], N[2]), dash-dotted curve: without R, and dashed
  curve: Born approximation according to Eq.\ 
  (\protect{\ref{eq:born}}).}
\label{fig:total}
\end{figure}

We start the discussion with the total cross section plotted in Fig.\ 
\ref{fig:total}.  The dotted curve corresponds to the result of I.
Adding MECs ($\Delta$[2] and N[2]) gives a slight increase of a few
percent in the maximum (dash-dotted curve). But by far much more
important are the other rescattering mechanisms. They reduce the cross
section significantly and shift the maximum to a slightly lower
position.  Furthermore, a comparison of the full to the dashed curve
clearly demonstrates that the perturbative treatment (Born
approximation) by replacing
\begin{equation}
  \label{eq:born}
  T_{\Delta\Delta}(z)\rightarrow V^{eff}_{\Delta\Delta}(z),
  \quad\quad
  T_{\Delta N}(z)\rightarrow V_{\Delta N}
\end{equation}
in the amplitudes of Eqs.\ (\ref{eq:rd}-\ref{eq:rn}), and
(\ref{eq:rmec}) is certainly insufficient.  It underestimates the full
dynamical effect by more than half and thus can provide a qualitative
description only.

As shown in Fig.\ 7 of I, the total cross section is essentially
determined by two matrix elements, namely the magnetic dipole
$M1(\gamma d)\rightarrow P_2(\pi d)$ and quadrupole $M2(\gamma
d)\rightarrow D_3(\pi d)$ transitions.  Their importance arises from
the coupling to the $^5S_2(N\Delta)$ and $^5P_3(N\Delta)$ partial
waves, respectively.  For a more complete list of channel couplings
relevant for $\gamma d\rightarrow\pi^0d$, we refer to Table 1 in I.
Unfortunately, the available experimental data for differential cross
sections do not allow a reliable determination of total cross sections
and thus a check of how well the present model describes the strength
of these two leading transitions.

\begin{figure}
  \centerline{\psfig{figure=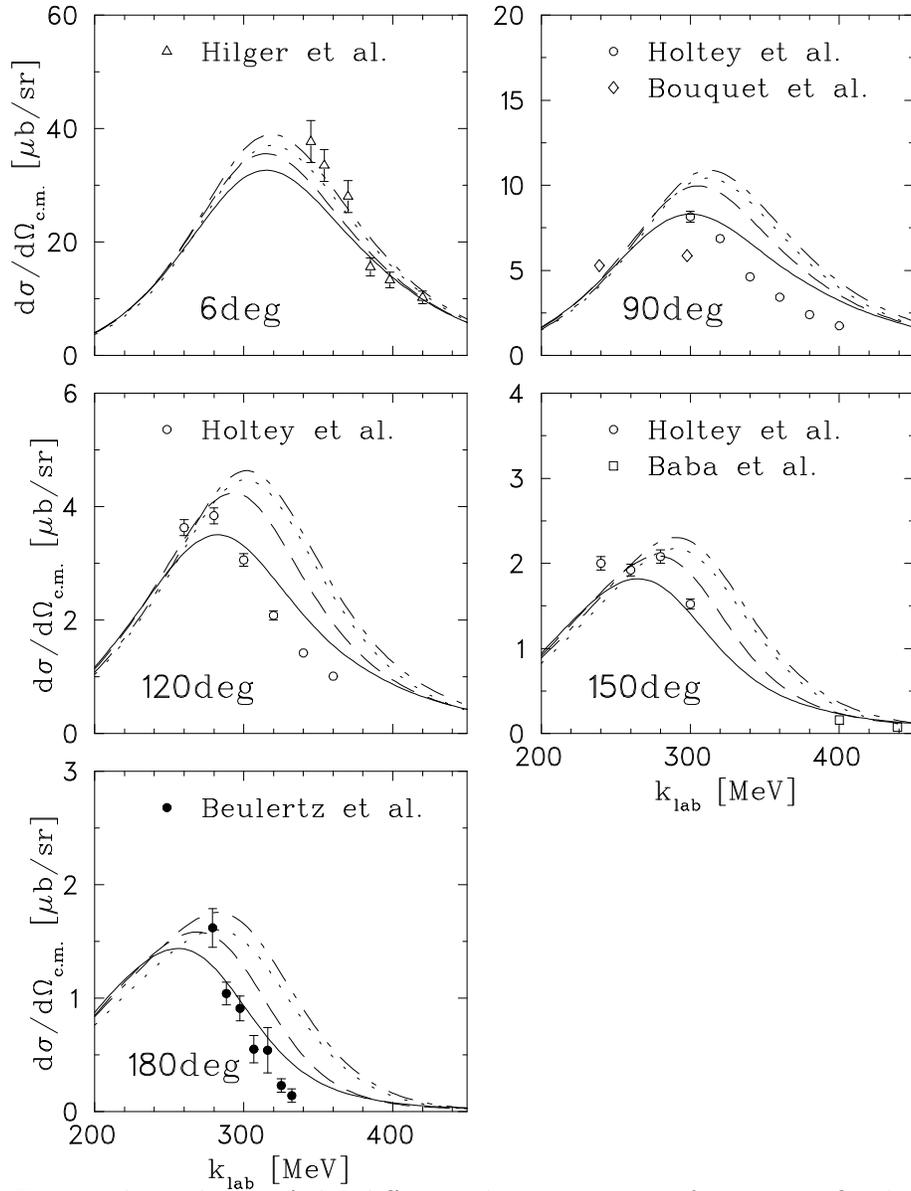,width=12cm,angle=0}}
\caption{
  Energy dependence of the differential cross sections for various
  fixed pion angles $\theta$.  Notation as in Fig.\ 
  \protect\ref{fig:total}.  Experimental data from
  \protect\cite{hilger,hol,bou,baba,beulertz}.}
\label{fig:dif}
\end{figure}

Differential cross sections for fixed pion angles, plotted in Fig.\ 
\ref{fig:dif}, show the same features.  For energies in the $\Delta$
region, MECs lead to slight increase but further pion rescattering
reduces the cross section at all angles.  Its influence strongly
grows with the pion angle.  This qualitatively agrees with what one
finds in pion deuteron elastic scattering.  Intuitively, one would
always expect that rescattering mechanisms become more important at
higher momentum transfers, i.e., for larger scattering angles at fixed
energy, since rescattering provides the possibility to share the
momentum transfer between the two nucleons.  In addition, Fig.\ 
\ref{fig:dif} demonstrates again that the Born approximation gives
only a qualitative description of rescattering effects.

As far as the comparison with experimental cross sections is
concerned, Fig.\ \ref{fig:dif} shows that the rescattering amplitudes
clearly remove most of the discrepancies between the data and our
results in I.  This conclusion is completely different from the one
drawn by Blaazer, Bakker and Boersma \cite{bbb}.  In their work, the
three-body rescattering contributions turned out to be fairly
unimportant, namely of the order of a few percent, and could in no way
resolve the overestimation of the PWIA cross section.  At present we
have no explanation for this striking deviation.  On the other hand,
the rescattering effects of Ref.\ \cite{ktb} agree qualitatively with
our results.  Moreover, the existing differences between these two
calculations can essentially be traced back to an insufficient
treatment of the charge exchange mechanism in Ref.\ \cite{ktb}.  This
aspect will now be discussed in some detail.

\subsection{The role of charge exchange amplitudes}
 
We have already mentioned that charge exchange amplitudes are
dropped within the KMT multiple scattering approach,
whereas in our model, the rescattering process includes the exchange
of both neutral and charged pions via the successive excitation and
decay of the $\Delta$ resonance as generated by the driving term
in (\ref{eq:ret}).  The effect of neglecting the charged pions
can directly be determined by splitting the isospin operator occurring
in (\ref{eq:ret}) into neutral plus charged pion contributions,
\begin{equation}
  \label{iso}
  \vec\tau_{\Delta N}(1) \cdot \vec\tau_{N\Delta}(2) 
  =\Omega(\pi^0)+\Omega(\pi^\pm),
\end{equation}
where
\begin{equation}
  \label{iso0}
  \Omega(\pi^0)=\tau_{\Delta N,0}(1)\,\,\tau_{N\Delta,0}(2).
\end{equation}
Now one immediately derives the ratio
\begin{equation}
  \label{isor}
  \frac{ \left\langle \left(\frac{1}{2}\frac{3}{2}\right)10
    \right|\Omega(\pi_0)\left|
      \left(\frac{3}{2}\frac{1}{2}\right)10\right\rangle } {
    \left\langle \left(\frac{1}{2}\frac{3}{2}\right)10
    \right|\Omega(\pi_0) + \Omega(\pi^\pm)\left|
      \left(\frac{3}{2}\frac{1}{2}\right)10\right\rangle } = 2,
\end{equation}
where the coupled two-particle isospin states have been denoted as
$|(t_1t_2)tm_t\rangle$.  The ratio (\ref{isor}) implies that the
neglect of the exchange of charged pions results in a significant
overestimation of the rescattering effect as far as it proceeds
through intermediate $\Delta$ excitation.  In a perturbative treatment
it would be just a factor two in the rescattering part of the reaction
amplitude.

\begin{figure}
  \centerline{\psfig{figure=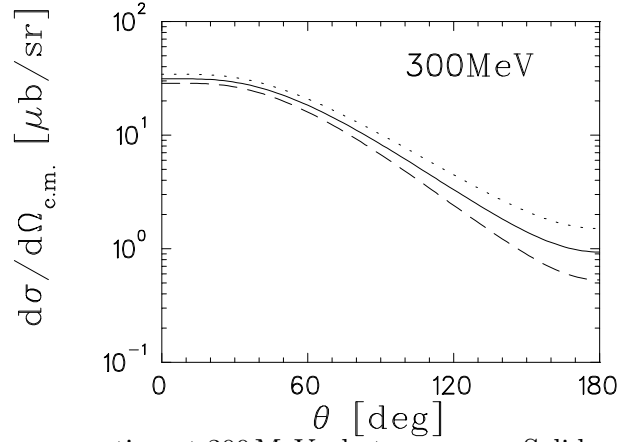,width=8cm,angle=0}}
\caption{
  Differential cross section at 300$\,$MeV photon energy.  Solid and
  dotted curves are as in Fig.\ \protect\ref{fig:total}.  Dashed
  curve: complete calculation without charged pion exchange in
  $V_{\Delta\Delta}^{eff}(z)$ as explained in the text.}
\label{fig:charge}
\end{figure}

In order to demonstrate the influence of charge exchange
quantitatively, we have performed a calculation for which the charged
pion exchange in the driving term $V_{\Delta\Delta}^{eff}(z)$ has been
switched off.  As a result the differential cross section at
300$\,$MeV is shown in Fig.\ \ref{fig:charge}.  For a more direct
comparison with the results of Kamalov, Tiator and Bennhold, we have
collected in Table \ref{table1} relative reduction factors of the
differential cross section due to pion rescattering as predicted by
the various calculations.  Apparently, our calculation without the
effect of the exchange of charged pions closely resembles the relative
effect of pion rescattering within the KMT formalism.  Moreover, the
results of Fig.\ \ref{fig:charge} and Table \ref{table1} show that the
claim of Ref.\ \cite{ktb}, namely that contributions from charge
exchange becomes small in the $\Delta$ resonance region, is not
correct.

\subsection{Polarization Observables}

\begin{figure}
  \centerline{\psfig{figure=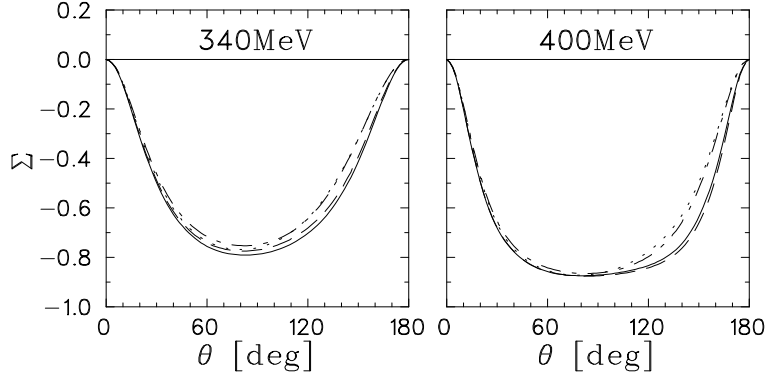,width=10cm,angle=0}}
\caption{
  Photon asymmetry $\Sigma$ at 340 and 400$\,$ MeV photon energy.
  Notation as in Fig.\ \protect\ref{fig:total}.}
\label{fig:sigma}
\end{figure}

In I we have defined all observables corresponding to a polarized
photon beam and/or an oriented deuteron target.  It turns out that
most of the polarization asymmetries are in general less influenced by
pion rescattering than the cross section.  As a typical example, we
show in Fig.\ \ref{fig:sigma} the photon asymmetry $\Sigma$.
Exceptions to this rule are those asymmetries which are not
constrained to vanish at 0 and 180$\,$deg, where one observes
significant rescattering effects at backward angles.  The only single
polarization observable of this type, the tensor target asymmetry
$T_{20}$, will now be discussed in greater detail.

\subsubsection{The tensor target asymmetry $T_{20}$}
\label{sec:t20}
 
For $\gamma d\rightarrow\pi^0 d$ at forward and backward angles, the
asymmetry $T_{20}$ allows to draw specific conclusions about details
of the reaction mechanism.  For this special kinematics, when the pion
momentum $\vec q$ is parallel or antiparallel to the photon momentum
$\vec k$, one has to deal with only two independent reaction
amplitudes instead of nine ones in general. This follows from the fact
that in the general representation of the full amplitude as
$T=\vec\epsilon_\lambda\cdot\vec J$, the current $\vec J$ has to be
constructed from the only available vectors $\hat k=\vec k/|\vec k|$
and $\vec S$, the total deuteron angular momentum.  Therefore, one has
for the spherical components of $\vec J$ at $\theta=0$ or $\pi$
\begin{equation}
  \label{eq:amp}
  J_\mu^{[1]} =
  A(W_{\gamma d},\theta=0,\pi)\, S_\mu^{[1]}
  +\sqrt{10}\, B(W_{\gamma d},\theta=0,\pi) \left[\left[
      S^{[1]}\times S^{[1]}\right]^{[2]} \times\left[ \hat
      k^{[1]}\times \hat k^{[1]}\right]^{[2]} \right]^{[1]}_\mu,
\end{equation}
where the square brackets denote the tensor coupling.  The coefficients
$A$ and $B$ define the two independent amplitudes. A complete set of
the $4+5$ amplitudes for the general case, where the first ones are
constructed in analogy to the four CGLN amplitudes for $\gamma
N\rightarrow \pi N$ and the latter five contain the tensor operator
$\left[S^{[1]}\times S^{[1]}\right]^{[2]}$, will be reported elsewhere
\cite{future} together with the expressions for all observables in
terms of these amplitudes.  In (\ref{eq:amp}), the normalization is
chosen such that the cross section is
\begin{equation}
  \label{eq:cross}
  \frac{d\sigma}{d\Omega}(\theta=0,\pi) \propto |A|^2+|B|^2.
\end{equation}
The asymmetry $T_{20}$ then reads
\begin{equation}
  \label{eq:t20}
  T_{20}(\theta=0,\pi) = -\frac{1}{2\sqrt{2}} \left(1+6\,
    \frac{\mbox{Re}(A^\ast B)}{|A|^2+|B|^2} \right).
\end{equation}
It is restricted to $-\sqrt{2}\le T_{20}\le 1/\sqrt{2}$.
Obviously, the deviation of $T_{20}$ from $-1/2\sqrt{2}$ is a
direct indication of a nonvanishing amplitude $B$.

It is this amplitude $B$ involving a tensor transition in deuteron spin
space which allows to draw some specific conclusions on the reaction
mechanism.  
They are based on the following observations:

({\it i}) In a deuteron model without $D$ wave, $\vec S$ is simply
given by the sum of the nucleon spins, $\vec
S=\frac{1}{2}\left(\vec\sigma(1)+\vec\sigma(2)\right)$, and thus the
tensor operator necessarily involves the spin of both nucleons since
$\left[S^{[1]}\times S^{[1]}\right]^{[2]}
=\frac{1}{2}\left[\sigma(1)\times\sigma(2)\right]^{[2]}$.  In this
case, a nonvanishing $B$ would unambiguously point to a two-nucleon pion
production mechanism.

({\it ii}) In a deuteron model with $D$ wave, there are also one-nucleon
pion production operators which can contribute to $B$.  However, local
one-nucleon operators, i.e., those which are independent from the Fermi
momentum of the active nucleon, cannot contribute to $B$.  It is the
last statement which makes the situation more conclusive here than for pion
deuteron elastic scattering, where already local one-nucleon scattering
operators in combination with the $D$ wave affect $T_{20}$ at extreme
angles (see Appendix C).

Thus in general, a nonvanishing $B$ amplitude points to either a
two-nucleon or to a nonlocal one-nucleon pion production mechanism,
where the latter one has to involve the deuteron $D$ wave.  It is not
surprising that two-nucleon and $D$ wave effects cannot be separated
experimentally since their relative weight is representation
dependent.  This has been demonstrated explicitly long ago by Friar
\cite{friar} who showed in a different context that the deuteron
quadrupole moment does not allow to fix the deuteron's $D$ wave
probability uniquely.

\begin{figure}
  \centerline{\psfig{figure=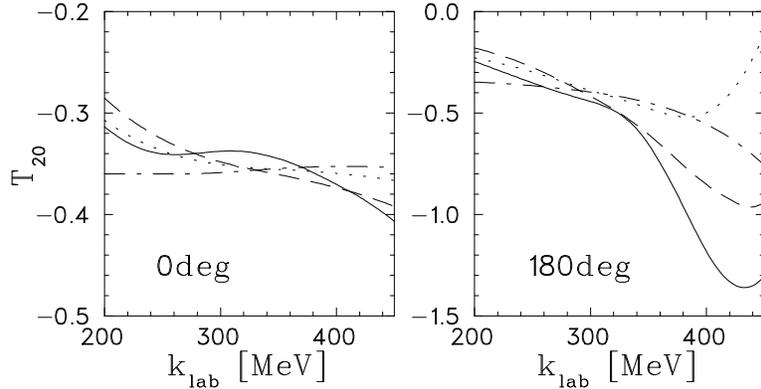,width=10cm,angle=0}}
\caption{
  Energy dependence of the tensor target asymmetry $T_{20}$ at 0 and
  180$\,$deg.  Dash-dotted curves include $\Delta$[1], NP[1], NC[1]
  and then consecutively added dTBA contribution NP[2], NC[2]
  (dotted), MEC contribution $\Delta$[2], N[2] (dashed), and further
  rescattering R (solid).}
\label{fig:t20}
\end{figure}

\begin{figure}
  \centerline{\psfig{figure=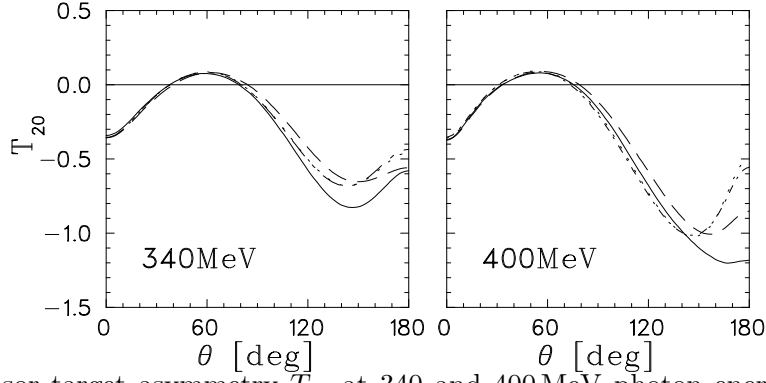,width=10cm,angle=0}}
\caption{
  Tensor target asymmetry $T_{20}$ at 340 and 400$\,$MeV photon
  energy. Notation as in Fig.\ \protect\ref{fig:t20}.}
\label{fig:xt20}
\end{figure}

Results for $T_{20}$ are plotted in Fig.\ \ref{fig:t20} for the angles
$\theta=0$ and 180$\,$deg as a function of the photon energy and in
Fig.\ \ref{fig:xt20} for the photon energies 340 and 400$\,$MeV as a
function of the angle $\theta$.  The dotted curves include the
amplitudes $\Delta$[1], NP[1] and NC[1] in the notation of Fig.\ 
\ref{fig:amplitudes} and the disconnected two-body amplitudes (dTBA)
NP[2] and NC[2] which have been discussed in I.  Particularly at lower
energies, they dominate the amplitude $B$ and lead to a significant
deviation from $-1/2\sqrt{2}$ as shown by a comparison with the
dash-dotted curves where the dTBA have been switched off.  MEC
contributions $\Delta$[2] and N[2], added in the dashed curves, show
sizeable effects at backward angles and higher energies.  Finally, the
solid curves include the rescattering term R. In Fig.\ \ref{fig:t20},
they introduce a characteristic energy dependence in the $\Delta$
region which in principle could be checked experimentally.

In comparison with Ref.\ \cite{ktb}, one notes a marked disagreement
for $T_{20}$ at backward angles.  In \cite{ktb}, $T_{20}(\theta=\pi)$
varies only between $-0.3$ and $-0.45$ for photon energies between 200
and 400$\,$MeV indicating a much smaller $B$ amplitude.  It shows that
the (deuteron) spin dependence of the rescattering process is very
different in both approaches. On the other hand, this result provides
a possibility to distinguish experimentally between both models.  With
respect to Refs.\ \cite{bbb} and \cite{pena} we can only compare with
results for 300$\,$MeV photon energy where we find good agreement.

We close this subsection with a remark concerning the reaction at
threshold. We first note that two nonvanishing amplitudes exist,
namely an electric dipole $E1(\gamma d)\rightarrow S_1(\pi d)$ and a
magnetic quadrupole $M2(\gamma d)\rightarrow S_1(\pi d)$.  They
coincide with $A$ and $B$, respectively, which become independent of
$\theta$ at threshold.  In view of a test of theoretical predicitions
for the $\gamma n\rightarrow\pi^0 n$ electric dipole at threshold,
e.g., from chiral perturbation theory, a measurement of $T_{20}$ in
the near threshold region may be very helpful. Because, according to
the above discussion, it could allow to clarify the role of
two-nucleon processes which hamper a direct extraction of the neutron
amplitude.  As a sideremark, we would like to point out that in a very
recent study of $\gamma d\rightarrow \pi^0 d$ in the framework of
heavy baryon chiral perturbation theory \cite{beane} the $M2$
transition has not been considered.

\subsubsection{The electric quadrupole excitation of the $\Delta$ resonance}

In I we have studied the sensitivity of the reaction to various
elementary pion production multipoles.  Within the PWIA the reaction
is sensitive to the coherent sum of the $\gamma p\rightarrow\pi^0p$
and $\gamma n\rightarrow\pi^0n$ amplitudes, usually denoted as
$A^+=\frac{1}{2}(A^{\pi^0p}+A^{\pi^0n})$.  Of particular interest is
the sensitivity to the $E_{1+}$ multipole which is closely related to
the electric quadrupole excitation of the $\Delta$ resonance.  Already
in I, we have pointed out that the vector target asymmetry $T_{11}$
provides an enhanced sensitivity to this quantity.  However, before
one can draw definite conclusions on the influence of the $E_{1+}^+$
multipole, the effect of the rescattering contribution has to be
investigated.

\begin{figure}
  \centerline{\psfig{figure=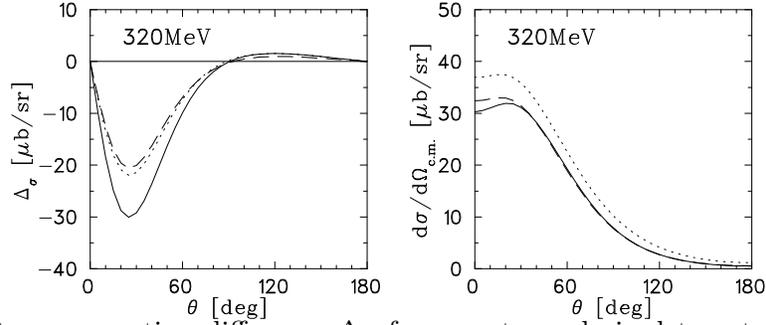,width=10cm,angle=90}}
\caption{
  Left: cross section difference $\Delta_\sigma$ for a vector
  polarized target, right: unpolarized cross section $d\sigma/d\Omega$
  at 320$\,$MeV photon energy. Solid curve: complete calculation with
  inclusion of the E2 excitation of the $\Delta$, dashed curve:
  without E2 contribution, and dotted curve: without rescattering
  contributions (R, $\Delta$[2], N[2]) and without E2.}
\label{fig:e2}
\end{figure}

Our results are summarized in Fig.\ \ref{fig:e2}.  The unpolarized
cross section $d\sigma/d\Omega$ and the cross section difference
$\Delta_\sigma=\frac{1}{p_z}(\frac{d\sigma^+}{d\Omega}
-\frac{d\sigma^-}{d\Omega})$ rather than the asymmetry $T_{11}$ are
plotted for 320$\,$MeV photon energy.
Denoting with 
$p_z$ and $p_{zz}$ the target vector and tensor
polarization parameters, respectively, we introduce the cross sections
\begin{equation}
\label{eq:deltasigma}
\frac{d\sigma^\pm}{d\Omega}= \frac{d\sigma}{d\Omega} \left[
  1\mp\frac{\sqrt{3}}{2}\, p_z\, T_{11} - \frac{1}{4}\, p_{zz}
  (\sqrt{3}\, T_{22}+\sqrt{2}\, T_{20}) \right],
\end{equation} 
with the deuteron orientation axis parallel and antiparallel to $\vec
k\times\vec q$, respectively.  In the unpolarized cross section, the
rescattering strongly masks the E2 excitation of the $\Delta$.  On the
contrary, in the difference $\Delta_\sigma$ the rescattering effect is
suppressed whereas the E2 leads to a change of the order of 50\% in
the most promising region between $\theta=20$ and 30$\,$deg. Assuming
$p_z\approx 1$, $p_{zz}\approx 0$, the cross sections actually to be
measured are of the order $d\sigma^+/d\Omega\approx 15\,\mu b/sr$ and
$d\sigma^-/d\Omega\approx 45\,\mu b/sr$.

\section{Summary and conclusion}

In this paper we have studied the influence of pion rescattering on
coherent pion photoproduction off the deuteron in the $\Delta$
resonance region.  The calculation is based on a dynamical model which
includes the coupling of N$\Delta$, NN$\pi$ and NN channels.  Our main
conclusions are as follows. Pion rescattering is significant and
reduces the cross section in the resonance region.  The strongest
influence appears at large momentum transfers (backward pion angles).
This means in particular with respect to a test of theoretical models
for pion production amplitudes on the neutron that one needs a
reliable description for the rescattering process.  Compared to
experimental data, we find that the sizeable discrepancies without
rescattering as reported in I, are largely reduced and that a
reasonable agreement with the data is achieved.

Furthermore, we have shown that a perturbative treatment of
rescattering, which is comparable to a double-scattering ansatz, can
only give a qualitative description of rescattering effects.  The
inclusion of charged pion exchange, i.e., the coupling to the
intermediate $\pi^+nn$ and $\pi^-pp$ break-up channels is by no means
unimportant as was claimed in \cite{ktb}.  Its neglection leads to a
sizeable overestimation of the rescattering effect.  This observation
could essentially explain the differences with respect to the relative
strength of the rescattering process between this work and the results
of Ref.\ \cite{ktb} which are based on the KMT multiple scattering
approach.  However, the discrepancies to the results of Ref.\ 
\cite{bbb} using a Faddeev ansatz remain unclear.

Most of the polarization observables are less sensitive to
rescattering than the cross section.  As an exception, we have
stressed the role of the tensor target asymmetry $T_{20}$ as a tool to
disentangle different reaction mechanisms.  Particularly, its
measurement at backward pion angles would be very helpful.
Incidentally, it would allow to distinguish between the present model
and the one of Ref.\ \cite{ktb}.  In Ref.\ \cite{WA1}, we have already
pointed out a remarkable sensitivity of the vector target asymmetry
$T_{11}$ on the E2 excitation of the $\Delta$ resonance.  Our analysis
here shows that rescattering effects do not seriously mask this
sensititvity.  In the most favorable case, such effects are suppressed
by about a factor 5 compared to the E2 effect.

\appendix
\section*{A}

Using a partial wave decomposition, Eq.\ (\ref{eq:cc}) reduces to a
one-dimensional integral equation which has to be solved numerically.
Its solution is hampered by the presence of logarithmic three-body
singularities which occur in $V_{\Delta\Delta}^{ret}(E^+)$ as a
consequence of the coupling to the open $\pi$NN channel.
In order to overcome this problem, we have iterated the
integral equation once.  Introducing for the moment being a more
convenient and obvious matrix notation we can write (\ref{eq:cc}) in
the form
\begin{equation}
  {\cal T}={\cal V} + {\cal {VGT}}.
\end{equation}
The once iterated equation 
${\cal T}={\cal V} + {\cal {VGV}} +
{\cal {VGVGT}}$ can be rewritten as an integral equation
for the difference ${\cal T}-{\cal V}$,
\begin{equation}
  {\cal T}-{\cal V}={\cal {VGV}} + {\cal {VGVGV}} + {\cal {VGVG}}
  \left({\cal T}-{\cal V}\right).
\label{eq:num}
\end{equation}
The advantage of (\ref{eq:num}) is that both, the kernel ${\cal {VGVG}}$
as well as the inhomogeneous term ${\cal {VGV}} + {\cal {VGVGV}}$, are
smooth quantities since the three-body singularities contained in
${\cal V}$ have been integrated over. Thus it can be solved
numerically by applying the matrix inversion technique after a
discretization in momentum space.

\section*{B}

The two-body current $\vec\jmath^{\,\,\pi}_{[2]\Delta N}$
has to be constructed consistent to the $\pi$ exchange in 
the transition potential which reads 
\begin{eqnarray}
 & V_{N\Delta}^{\pi} = &
 -\frac{f_{\pi N N}f_{\pi N \Delta}}{m_{\pi}^2}
 \vec{\tau}_{NN}(1) \cdot \vec{\tau}_{N\Delta}(2) \\
 &&\int \frac{d^3q}{(2\pi)^3} \:e^{i\vec q\cdot\vec r}\: \frac
 {\vec{\sigma}_{NN}(1) \cdot \vec{q} \,
 \vec{\sigma}_{N\Delta}(2) \cdot \vec{q}}
 {\vec{q}\,^2+m_{\pi}^2}
 F_{\pi N N}(\vec{q}\,^2) F_{\pi N\Delta}(\vec{q}\,^2)
 \,+\,(1 \leftrightarrow 2),\nonumber
\end{eqnarray}
where $\vec r$ is the relative coordinate of the two baryons.
In the present version of our model (without $\rho$ exchange in 
$V_{N\Delta}$), the form factors have been taken in monopole form
\begin{equation}
  F_{\pi N \Delta}(\vec{q}\,^2)=F_{\pi N N}(\vec{q}\,^2)
  = \frac{\Lambda_\pi^2-m_\pi^2}{\Lambda_\pi^2+\vec q^{\,2}},
\end{equation}
where $\Lambda_\pi=700\,$MeV has been determined by fitting the NN
scattering data in the $^1D_2$ channel (see \cite{wilhelm}) using
$f_{\pi NN}^2/4\pi=0.08$ and $f_{\pi N\Delta}^2/4\pi=0.35$.  The MEC
can be written as
\begin{equation}
 \vec\jmath^{\,\,\pi}_{[2]\Delta N}(\vec{k})=
 ie \left(\vec{\tau}_{\Delta N}(1)\times\vec{\tau}_{NN}(2)\right)_0
 \frac{f_{\pi N\Delta}f_{\pi NN}}{m_{\pi}^2}
 \int \frac{d^3q}{(2\pi)^3} \,e^{i\vec q\cdot\vec r}
 \sum_{x=k_\pm,\pi,v} \vec{V}^{x}
 \Gamma^x
 +(1\leftrightarrow 2),
\end{equation}
where the
vertex and propagator structures of the contact ($k_\pm$), 
pion-in-flight ($\pi$), and vertex regularization ($v$) currents 
are given by
\begin{eqnarray}
 &&
 \vec{V}^{k_-}=\vec{\sigma}_{\Delta N}(1) \vec{\sigma}_{NN}(2)\!\cdot\!
 \left(\vec{q}-\vec{k}/2\right),\quad\quad\\
 &&
 \vec{V}^{k_+}=\vec{\sigma}_{NN}(2) \vec{\sigma}_{\Delta N}(1)\!\cdot\!
 \left(\vec{q}+\vec{k}/2\right),\quad\quad\\
 &&
 \vec{V}^{\pi}=\vec{V}^{v}=\vec{q}\,
 \vec{\sigma}_{\Delta N}(1)\!\cdot\!\left(\vec{q}+\vec{k}/2\right)
 \vec{\sigma}_{NN}(2)\!\cdot\!\left(\vec{q}-\vec{k}/2\right),\\
 &&
 \Gamma^{k_{\pm}}=
 \frac{F_{\pi N\Delta\pm}F_{\pi NN\pm}} {\omega_{\pm}^2},\quad\quad\\
 &&
 \Gamma^{\pi}=-2\,
 \frac{F_{\pi N\Delta +}F_{\pi NN-}} {\omega_+^2\omega_-^2},\quad\quad\\
 &&
 \Gamma^{v}=
 2\,\frac{F_{\pi N\Delta +}-F_{\pi N\Delta -}} {\omega_+^2-\omega_-^2}
 \frac{F_{\pi NN -}} {\omega_-^2} +
 2\,\frac{F_{\pi NN +}-F_{\pi NN -}} {\omega_+^2-\omega_-^2}
 \frac{F_{\pi N\Delta +}} {\omega_+^2},\quad\quad
\end{eqnarray}
with $\omega^2=\vec q^{\,2}+m_\pi^2$ and using the notation
$f_{\pm}=f(\vec q\pm\vec k/2)$ for any function $f(\vec q\,)$.  The
two-nucleon currents $\vec\jmath^{\,\pi}_{[2]NN}$ and
$\vec\jmath^{\,\rho}_{[2]NN}$ consistent to the OBEPR potential may be
found in \cite{kms}.

\section*{C}

First we show that local one-nucleon operators cannot contribute to
the tensor transition amplitude $B$ defined in (\ref{eq:amp}).  In
general, all transitions which contribute to a coherent deuteron
reaction can be classified as scalar, vector, and tensor (rank 2)
transitions with respect to the deuteron angular momentum space. In
order to identify them for a given transition operator ${\cal O}$, we
start by writing the deuteron wave function with magnetic quantum
number $m$ as
\begin{equation}
  \label{eq:dwf}
  \psi_m(\vec r\,)=\frac{1}{\sqrt{4\pi}}\frac{1}{r}
  \left(u_0(r)+\frac{1}{\sqrt{8}}S_{12}\,u_2(r)\right)
  \left|\left(\frac{1}{2}\frac{1}{2}\right)1m\right \rangle,
\end{equation}
where $S_{12}=3\vec\sigma(1)\cdot\hat r\,\,\vec\sigma(2)\cdot\hat r
-\vec\sigma(1)\cdot\vec\sigma(2)$.
For a $S\leftrightarrow S$, $S\leftrightarrow D$, or $D\leftrightarrow D$
transition one has to consider with respect to the spin degrees of freedom 
the effective operators
\begin{eqnarray}
  \label{eq:eff}
  {\cal O}_{SS} &=& {\cal O},\\
  {\cal O}_{SD} &=& S_{12}{\cal O}+{\cal O}S_{12},\\
  {\cal O}_{DD} &=& S_{12}{\cal O}S_{12},
\end{eqnarray}
respectively.  Note, that ${\cal O}$ itself may depend on
$\vec\sigma(1)$ and $\vec\sigma(2)$.  Using the following equivalences
valid within the space
$\left\{|(\frac{1}{2}\frac{1}{2})1m\rangle\right\}$
\begin{eqnarray}
  \vec\sigma(1)\cdot\vec\sigma(2) & \simeq & 1,\\
  \vec\sigma(1)+\vec\sigma(2) & \simeq & 2\vec S,\\
  \vec\sigma(1)-\vec\sigma(2) & \simeq & 0,\\
  \vec\sigma(1)\times\vec\sigma(2) & \simeq & 0,\\
  \left[\sigma^{[1]}(1)\times\sigma^{[1]}(2)\right]^{[2]} 
  & \simeq & 2 S^{[2]},\\ 
\end{eqnarray}
where $\vec S$ is the total deuteron angular momentum 
and $S^{[2]}=\left[S^{[1]}\times S^{[1]}\right]^{[2]}$,
one can rewrite the effective operators ${\cal
  O}_T$ ($T=SS$, $SD$, $DD$) as
\begin{equation}
  \label{eq:ot}
  {\cal O}_T = {\cal O}_T^{[0]} + \left[ {\cal O}_T^{[1]}\times
    S^{[1]}\right]^{[0]} + \left[ {\cal O}_T^{[2]}\times S^{[2]}
  \right]^{[0]},
\end{equation}
where ${\cal O}_T^{[L]}$ does not contain any spin operators.

In our case, namely for $\gamma d\rightarrow\pi^0 d$ at 0 or
180$\,$deg, any local one-nucleon operator has the structure
$\vec\sigma\cdot\vec v$ with a vector $\vec v$ and $\left[\vec v,\vec
  r\,\right]=0$, since no local pseudovector apart from $\vec\sigma$
is available (a nonlocal one would be given by $\vec p\times\hat k$
where $\vec p$ is the nucleon momentum).  The above statement then
follows using the identities
\begin{eqnarray}
  \label{eq:id1}
  S_{12}\,\vec\sigma(1/2)\cdot\vec v + \vec\sigma(1/2)\cdot\vec
  v\,S_{12} &=& 6\vec v\cdot\hat r\,\vec\sigma(2/1)\cdot\hat r - 2
  \vec\sigma(2/1)\cdot\vec v,\\ S_{12}\,\vec\sigma(1/2)\cdot\vec v\,
  S_{12} &=& \left(\vec\sigma(1)+\vec\sigma(2)\right)\cdot \left(
    6\vec v\cdot\hat r\,\hat r -4\vec v\right),
\end{eqnarray}
which do not contain a tensor transition on their right hand sides.

Next, we point out the difference of $\gamma d\rightarrow\pi^0 d$ to
the case of the $\pi d\rightarrow\pi d$ reaction at 0 or 180$\,$deg.
For the latter reaction, the analogous relation of Eq.\ (\ref{eq:amp})
reads
\begin{equation}
  T_{\pi d} = A(W_{\pi d},\theta=0,\pi) +\frac{3}{2}\sqrt{10}\, B(W_{\pi
    d},\theta=0,\pi) \left[S^{[2]} \times\left[ \hat k^{[1]}\times
      \hat k^{[1]}\right]^{[2]} \right]^{[0]}_0,
\end{equation}
with corresponding cross section as in (\ref{eq:cross}). The asymmetry
is given by
\begin{equation}
  T_{20}(\theta=0,\pi) = -\frac{1}{\sqrt{2}}
  \frac{|B|^2-2\sqrt{2}\,\mbox{Re}(A^\ast B)}{|A|^2+|B|^2}.
\end{equation}
In the case of pion scattering, one deals with scalar transition operators
and thus already local one-nucleon operators involving
$S\leftrightarrow D$ or $D\leftrightarrow D$ transitions contribute to
the amplitude $B$ which can be observed by means of the deviation from
$T_{20}=0$ at extreme angles.


\begin{table}
\caption{
  Relative reduction factors of the differential cross section at
  300$\,$MeV photon energy for various pion angles due to pion
  rescattering.  The second and third column correspond to the ratios
  of the dotted to the solid and to the dashed curves in Fig.\ 
  \protect\ref{fig:charge}, respectively.  The last column contains
  the ratios as read off from the dashed and solid curves in Fig.\ 4
  of Ref.\ \protect\cite{ktb}.}
\label{table1}
\begin{tabular}{rrrr}
  $\theta$ (deg) & this model with $\pi^{\pm}$ & this model
  without $\pi^{\pm}$ & model of Ref.\ \protect\cite{ktb} \\ 
  \tableline
    0 & 1.10 & 1.20 & 1.3 \\
   90 & 1.22 & 1.51 & 1.5 \\ 
  180 & 1.61 & 2.83 & 3.0 \\
\end{tabular}
\end{table}
\end{document}